\documentclass[aps,prb,twocolumn,floatfix]{revtex4-1}
\usepackage{graphicx,epsfig,array,color}
\usepackage{color,graphicx,amssymb,amsmath,bm}
\usepackage[utf8]{inputenc}
\usepackage[colorlinks=true,linktoc=page,linkcolor=magenta,
citecolor=cyan]{hyperref}
\usepackage{braket}

\usepackage{mathtools}

\def\<{\langle}
\def\>{\rangle}

\def\nn{\nonumber}

\def\beq{\begin{equation}}
\def\eeq{\end{equation}}

\newcommand{\bea}{\begin{eqnarray}}
\newcommand{\eea}{\end{eqnarray}}

\def\ket#1{| \,#1\, \rangle}
\def\bra#1{\langle \,#1\, |}

\usepackage{ragged2e}

\begin{document} 
\title{Essential role of  quantum speed limit in violation of Leggett-Garg inequality across a PT-transition}


\newcommand*{\affaddr}[1]{#1} 
\newcommand*{\affmark}[1][*]{\textsuperscript{#1}}


\author{Anant V. Varma \affmark[1] }
\email{anantvijay.cct@gmail.com}

\author{Jacob E. Muldoon \affmark[2]}
\email{jemuldoo@iupui.edu}

\author{Sourav Paul \affmark[1]}
\email{sp20rs034@iiserkol.ac.in}

\author{Yogesh N. Joglekar \affmark[2]}
\email{yojoglek@iupui.edu}

\author{Sourin Das \affmark[1]}
\email{sourin@iiserkol.ac.in; sdas.du@gmail.com \\}

\affiliation{
\affaddr{\affmark[1] Indian Institute of Science Education and Research Kolkata, Mohanpur, Nadia 741246, West Bengal, India.}\\
\affaddr{\affmark[2] Department of Physics, Indiana University Purdue University Indianapolis (IUPUI), Indianapolis, Indiana 46202,  USA.}}

\begin{abstract}
We study Leggett-Garg inequality (LGI) of a two level system (TLS) undergoing non-Hermitian dynamics governed by  a non-linear Bloch equation (derived in J. Phys. A: Math. Theor. {\bf 54}, 115301 (2021)) across a PT-transition. 
We  present an algebraic identification of  the parameter space for the maximum violation of LGI (in particular $K_{3}$). In the PT-symmetric regime the maximum allowed value for $K_{3}$ is always found to be greater than the quantum bound (L\"{u}ders bound) of $3/2$ but it does not reach the algebraic maximum of $K_{3}=3$ in general. However, in the limit where PT-symmetry breaking parameter approaches  the exceptional point from the PT-symmetric side, $K_{3}$ is found to asymptotically approach its algebraic maximum of 3. In contrast, the maximum value of $K_{3}$ always reaches its algebraic maximum in the PT-broken phase $i.e.$ $K_{3}\rightarrow 3$. We find that (i) the speed of evolution (SOE) must reach its maximum value
 (in the parameter space of initial state and the time interval between successive measurements) to facilitate the value of $K_{3} \rightarrow 3$, (ii) 
together with the constraint that its minimum value must run into SOE equals to zero during the evolution of the state. In fact we show that the minimum speed of evolution can serve as an order parameter which is finite on the PT-symmetric side and identically zero on the PT-broken side.
Finally, we discuss a possible experimental realization of this dynamics by quantum measurement followed by post-selection procedure in a three level atom coupled to cavity mode undergoing a Lindbladian dynamics.   
\end{abstract}

\maketitle
\section{Introduction}{\label{sec:I}}

Violation of Leggett-Garg inequalities are well-studied markers of quantum dynamics  \citep{Bell1964, Peres1999, Mahler1993, Kim2006, Budroni2013} once complemented with criterion like no-signalling in time. These inequalities have been exploited to  test quantum mechanics at macroscopic scales \cite{LG1985,Leggett2008,Legget2002}. LGIs are based  on our intuition of the classical world defined in terms of two postulates namely (a) macroscopic realism and (b) non-invasive measurability ~\cite{LGIreview2014}. There are  many different variants of these inequalities and the simplest three-time measurement scenario can be expressed as: $-3 \leq K_{3}=C_{12}+C_{23}-C_{13}\leq 1$, where $ C_{ij}$ represent the two time-correlations, where $C_{ij} = P_{ij}(\uparrow,\uparrow) + P_{ij}(\downarrow,\downarrow)- P_{ij}(\uparrow,\downarrow)- P_{ij}(\downarrow,\uparrow)$, where  $P_{ij}$s represents the joint probability of the  two outcomes $(\uparrow,\downarrow)$ of the dichotomic observable where measurements are performed at time $t_i$ and  $t_j$.  In case of Hermitian dynamics defined by Hamiltonian $H$ typically and an arbitrary initial state  $\ket{\psi}$ two-time correlation function $C_{ij}$ is given by the following anti-commutator form \cite{Fritz2010}: $C_{ij} \ = (1/2) \ \bra{\psi}    \{ Q(t_{i}) , Q(t_{j}) \}    \ket{\psi}$, where $Q(t_{i})= e^{i H t_{i}}  Q  e^{-i H t_{i}}$ and $Q(t_{j})= e^{i H t_{j}}  Q  e^{-i H t_{j}}$ are time evolved versions of dichotomic observable $Q$ in the Heisenberg picture. The maximum value of $K_3$ for an $N$ level system is $3/2$ and is termed as L\"{u}ders bound ~\cite{Budroni2013}. Violation of this bound for an $N$  level quantum system, where $N>2$ is possible provided further degeneracy breaking measurements are performed \cite{Emary2014} but violation of L\"{u}ders bound for $N=2$ i.e. a two level system (TLS) is impossible within the unitary dynamics.  \\
 
\begin{widetext}

\begin{figure}[htb!]
    \centering
    \vspace{1em}
\includegraphics[width=5.25cm,height=5.5cm]{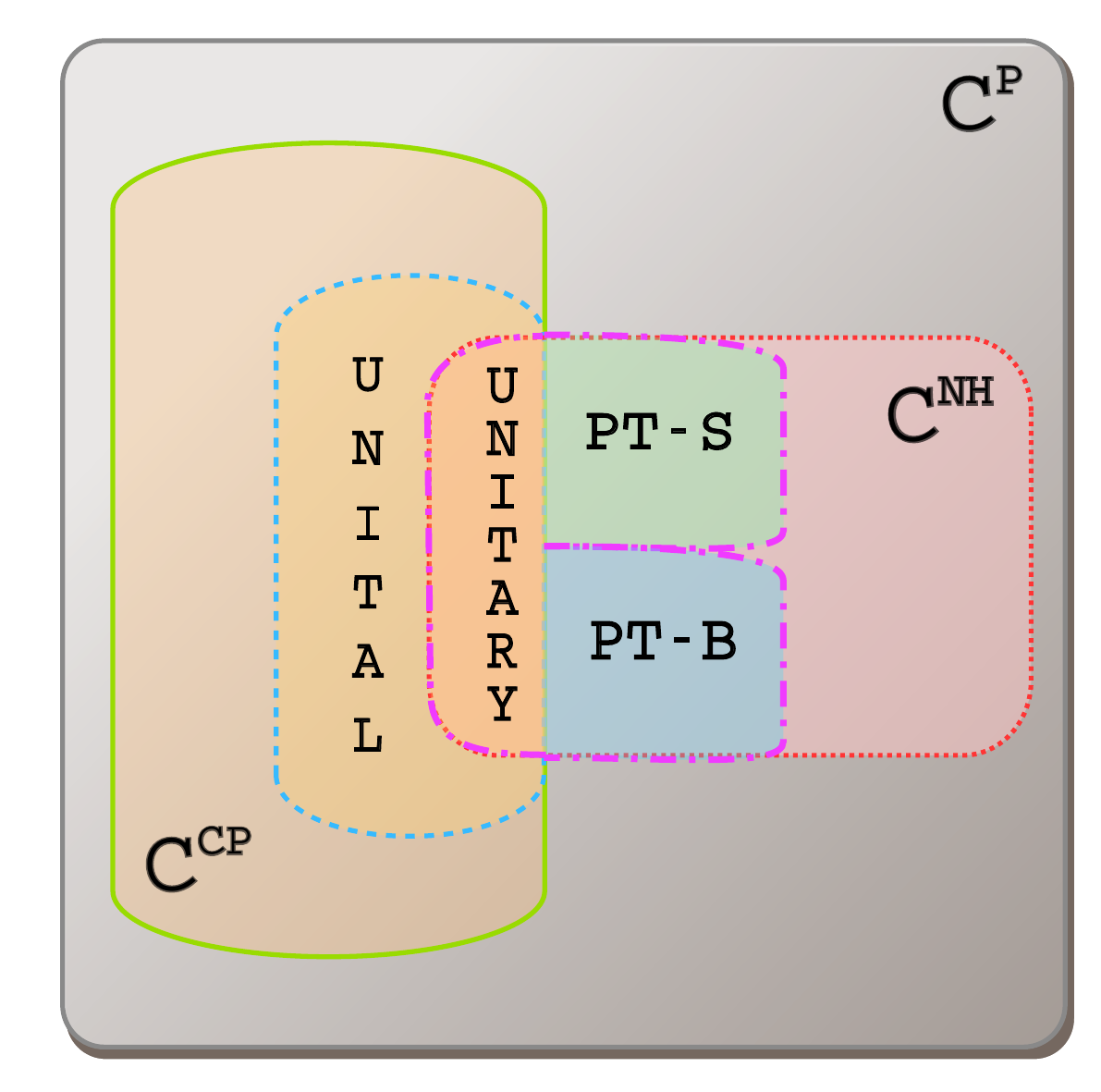} \hspace{-0.85em}
\includegraphics[width=7.5cm,height=6.0cm]{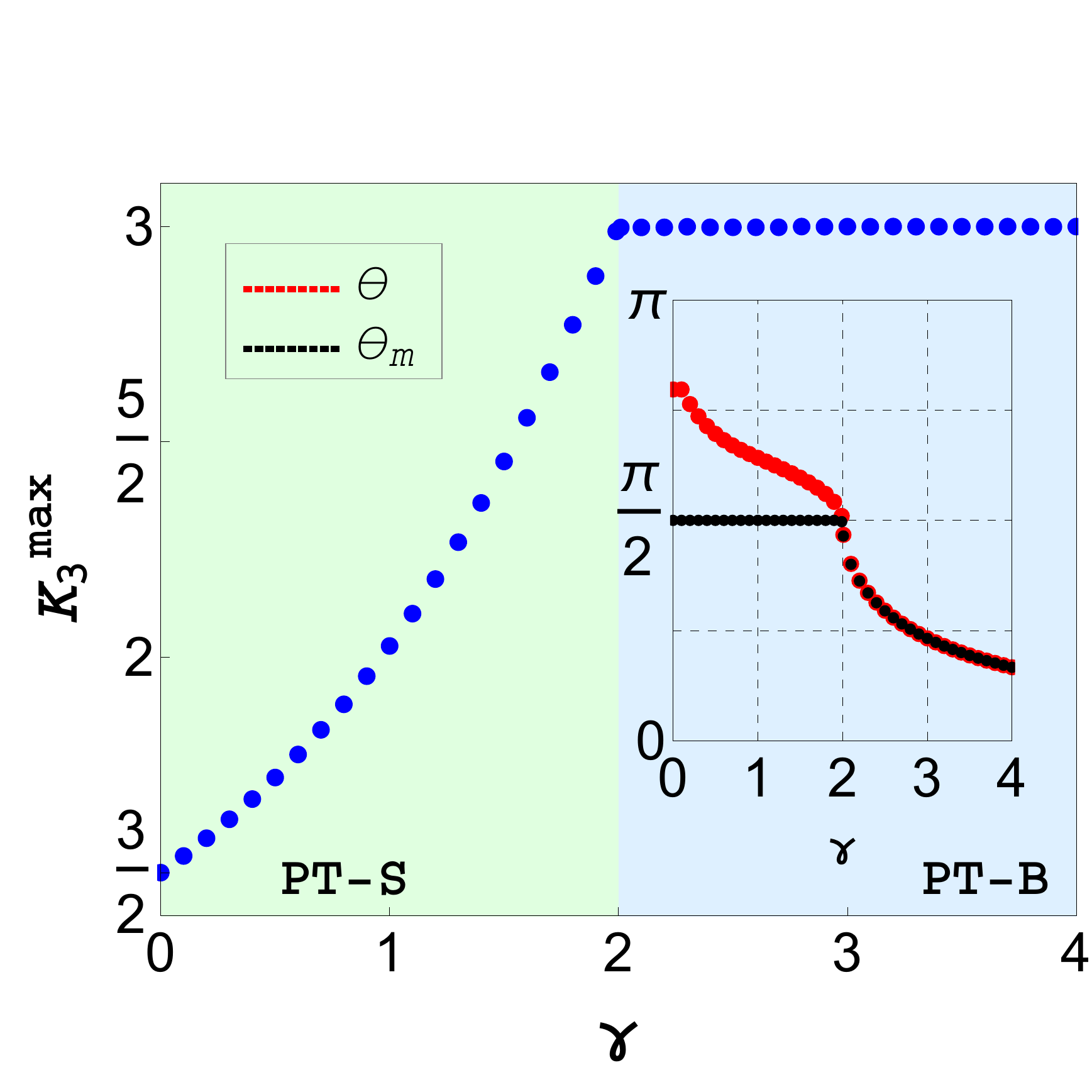}  \hspace{0.75em} 
\includegraphics[width=3.0cm,height=5.5cm]{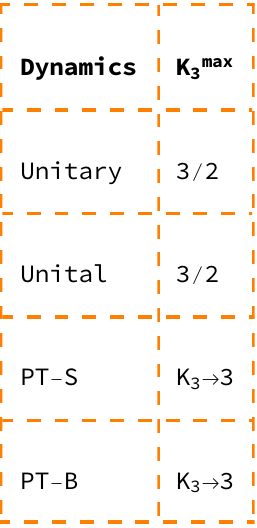}
\caption{(Left) Graphical depiction of different dynamics possible for a TLS. Here $C^{P}$ represents complete set of dynamics which are positive and trace preserving, while $C^{CP}$ corresponds to subset of $C^{P}$ which are completely positive (CP) and trace preserving. Non-Hermitian dynamics comprise that subset of $C^{P}$, which is $CP$ only when dynamics is unitary and represented by $C^{NH}$. PT-symmetric and PT-broken phases are represented by $PT-S$ and $PT-B$ respectively. (Middle) Maximum values of $K_{3}$ denoted as $K_{3}^{max}$ are plotted w.r.t. $\gamma/J$, where $J=1$ without any loss of generality. In the inset, are the corresponding values of initial state parameter $\theta$, and measurement operator parameter  $\theta_{m}$, where initial state $\ket{\psi}= [\ \cos \frac{\theta}{2} e^{i \phi}, \sin \frac{\theta}{2} ]^{T}\ $ and measurement operator is $\vec{n} \cdot \sigma$ with $\vec{n}= [\sin \theta_{m} \ \cos \phi_{m},\sin \theta_{m} \ \sin \phi_{m}, \cos \theta_{m}  ]$. Also, the values of $\phi = 3 \pi/2$ and $\phi_{m}=\pi/2$ for all the optimized values of $K_{3}$.(Right) Table showing maximum values of $K_{3}$ for different dynamics.}
\label{fig1}
\end{figure}

\end{widetext}

 However, any quantum system in a natural setting is expected to have finite coupling with its environment degrees of freedom leading to  non-unitary dynamics. All dynamics which are completely positive (CP) and trace preserving are considered as valid dynamics in quantum mechanics \cite{sudarshan, plenio1,lock,vedral,lupo,lian}. Unitary dynamics is the most common example of such CP dynamics. Rest of the set of CP dynamics is comprised of  unital and non-unital dynamics. It is known by now, in case of both the unitary and unital dynamics that upper bound on the LG parameter $K_{3}$ is L\"{u}ders bound \cite{LGIreview2014,sayan}. However,  set of dynamics  outside the sub-set of CP dynamics, which is only positive and trace preserving has not been explored much earlier. In this article we focus on  on such set of dynamics popularly referred  as non-Hermitian dynamics $i.e.$  $C^{NH}$ (see FIG. \ref{fig1} (Left)).
Recently it has been shown that $K_{3}$ can take values upto its algebraic bound  in case of a TLS undergoing non-Hermitian dynamics (PT-symmetric with fixed eigenspectrum) \cite{Anant,Usha} and close to the exceptional point in the PT-symmetric phase \cite{Pan}. Whereas we explore the non-Hermitian dynamics of a TLS across the exceptional point separating the PT-symmetric and PT-broken phases. We find that when the dynamics is in PT-symmetric phase it is impossible to find values of $K_{3}$ approaching its algebraic maximum i.e. 3, except close to the exceptional point only. More interestingly we find that it is always possible to approach the maxima $K_{3}\rightarrow 3$  in the PT-broken phase. We therefore explore and present the understanding of how the  initial state, measurement operator and SOE of the state conspire together leading to such extreme temporal correlations for PT-broken phase as well as close to the exceptional point in the PT-symmetric phase. We find that change in the SOE and the time interval dependence of the evolution, which are impossible in unitary dynamics are responsible for such extreme correlations when measurement is chosen  appropriately. The maximum SOE must reach its maximum value in the optimized parameter space, which is comprised of parameters of initial state, and time interval between measurements. Moreover, the minimum value of SOE must become zero during the evolution of the state between the two successive measurements for algebraic maximum of $K_{3} \rightarrow 3$ . Interestingly, we find that this minimum SOE can be treated as order parameter to characterize the PT-transition as its value is finite on the PT-symmetric side and identically zero on the PT-broken side..

The remainder of this article is organized as follows. In section ~\ref{secII}  we revisit solution of the dynamical equation representing non-Hermitian dynamics in terms of spin components of the Bloch vector  and find their relation with the SOE of the state. In section ~\ref{secIII} we find maximum values of LG parameter $K_{3}$ in both PT-symmetric and broken regime optimized in the full parameter space. We then present the dynamical process and its relation with the SOE, resulting in the algebraic maximum $i.e.$ $K_{3}=3$. Section ~\ref{secIV} is devoted to the realization of the non-Hermitian dynamics of TLS by exploiting a three level atom coupled to an environment undergoing Lindbladian dynamics. Finally, we dedicate section \ref{secV} on discussion and conclude in the section \ref{secVI} .

\section{Dynamical solution and SOE \label{secII}}

 If any arbitrary initial state $ \rho_{0}$ of a TLS is evolved through a generic non-Hermitian Hamiltonian $ H_{PT} \ = (\vec{A} - i \vec{B} ). \vec{ \sigma}$, then the governing dynamical equation is given by \cite{Brody} :

\begin{equation}
    \frac{d \rho_{t}}{dt} =  -i \left [ \vec{A}\cdot \vec \sigma,\rho_{t} \right ] -\left \{ \vec{B}\cdot \vec \sigma,\rho_{t}  \right \} + 2 \ tr( \rho_{t}  \vec{B}\cdot \vec \sigma) \rho_{t}.
    \label{0}
\end{equation}

The formal solution of the eq.(\ref{0}) is given by:

\begin{equation}
    \rho_{t} =(1/2)~  \mathbb{I} + \vec{S}(t) \cdot \vec \sigma = \frac{e^{- i H_{PT} t} \rho_{0} e^{i H_{PT}^{\dagger} t}}{tr (e^{- i H_{PT} t} \rho_{0} e^{i H_{PT}^{\dagger} t})}. 
    \label{1}
\end{equation}

It should be noticed that the two eigenstates of the Hamiltonian $H_{PT}$ are the fixed points of this dynamics. Assuming $\ket{\pm}$ are the right eigenstates of $H_{PT}$ with eigenvalues $e_{\pm}$, then $e^{- i H_{PT} t} \rho_{0} e^{i H_{PT}^{\dagger} t} = e^{- i H_{PT} t} \ (\ket{\pm} \bra{\pm}) \  e^{i H_{PT}^{\dagger} t}= e^{-i (e_{\pm}- e_{\pm}^{*}) t} \ (\ket{\pm} \bra{\pm})$ and $tr (e^{- i H_{PT} t} \rho_{0} e^{i H_{PT}^{\dagger} t}) =e^{-i (e_{\pm}- e_{\pm}^{*}) t} $ implying the fixed point behaviour of the two eigenstates.
The corresponding Bloch like equation controlling the time evolution of the Bloch vector $\vec{S}$ corresponding to eq.(\ref{0}) is given by: 

\begin{equation}
\frac{d\vec{S}(t)}{dt}  = 2 \ \vec{A}\times \vec{S}(t)  -  \vec{B} + 4 \  {\{} \vec{B} \cdot \vec{S}(t)  {\}} \ \vec{S}(t)~.
\label{2}
\end{equation}

To obtain analytical solutions of eq. (\ref{2}), we transform the coordinate system and work in the Cartesian coordinate system defined by the unit vectors $\hat{A}$, $\hat{B}$ and $\hat{n}= \hat{A}\times \hat{B}$. Rewriting eq. (\ref{2}) in its component form, we obtain
\begin{eqnarray}
\frac{dS_{A}(t)}{dt} &=&  4 \left |B  \right | S_{A} S_{B}~,
\label{3} 
\end{eqnarray}
\begin{eqnarray}
\frac{dS_{B}(t)}{dt}  &=& -2 S_{n}  \left |A  \right | - \left |B  \right | + 4 \left |B  \right | S_{B}^{2}~,
\label{4} 
\end{eqnarray}
\begin{eqnarray}
\frac{dS_{n}(t)}{dt}  &=& 2 S_{B}\left |A  \right |  + 4 \left |B  \right | S_{B}  S_{n}~, 
\label{5} 
\end{eqnarray}
where  $S_{A}(t)= \vec{S}(t) . \hat{A}, ~ S_{B}(t)= \vec{S}(t) . \hat{B},~ S_{n}(t) = \vec{S}(t).\hat{n}$ \& $\hat{n}= \hat{A} \times \hat{B}$. We can now define the SOE of the Bloch vector $S$. Using the definition of  SOE by Anandan and Aharonov~\cite{Anandan1990}, we can write  by expanding the term $\left |\braket{\psi(t)|\psi(t + \delta t)}\right |^{2} $ as: $\left |\braket{\psi(t)|\psi(t + \delta t)}\right |^{2} = 1- v^{2}(t) \delta t^{2} + O(\delta t^{3})$ ~, where $v(t) = \sqrt{ \  v_{1}^{2}(t) + v_{2}^{2}(t) + v_{3}^{2}(t)}$  is identified as the SOE. Here $v^{2}_{1}=(\Delta \vec{A} \cdot \vec \sigma)^{2}$, $v^{2}_{2} = (\Delta \vec{B} \cdot \vec \sigma)^{2} $ and $v^{2}_{3} = - i \left \langle [\vec{A} \cdot \vec \sigma,\vec{B} \cdot \vec \sigma]_{C} \right \rangle$ \cite{Brody, Anant}.  
Numerical  scan  shows  that  the  maximum  values  of $K_{3}$ lies in the $\hat{B}- \hat{n}$ plane.  Therefore, we now work in the $S_{A}=0$ subspace. In this subspace $i.e.$ $S_{a}(t)=0$. We find that all the we can write down the solutions of eq. (\ref{4}-\ref{5}) in a compact form as:

\begin{equation}
S_{B}(t)=  - \frac{1}{2} ~\sqrt{\frac{(A^2-B^2) \ \sin^2 [ 2 \sqrt{A^2-B^2} \ (t+C) ] }{(  \ A-B \cos [ 2 \sqrt{A^2-B^2} \ (t+C) ]  \ )^2}}~,
\label{6} 
\end{equation}
\begin{equation}
 S_{n}(t) = -\frac{1}{2} \frac{B-A \cos [ 2 \sqrt{A^2-B^2} \ (t+ C) ] }{A- B \cos [ 2 \sqrt{A^2-B^2} \ (t+C) ] }~,
 \label{7} 
\end{equation}
where constant $C$ is determined by initial values of $S_{B}$ and $S_{n}$ at time $t=0$. Since these solutions represent the evolution on the geodesic in the plane $\hat{B}-\hat{n}$, and there are two fixed points on the the geodesic. Therefore, it is important to ask which of the two paths on the geodesic separated by two fixed points the state follows. In order to specify that we Taylor expand the solution $S_{B}(t)$ in time upto first order. We find that the sign of the coefficient of first order term which is proportional to $ S_{n}(0) \ ( \  A+2 B S_{n}(0)  \ )$ decides the path followed by the initial state.

It should be noted that $\sqrt{A^2-B^2}>0$ $i.e.$  $\gamma/J<2$ then we are in PT-symmetric regime and when $\sqrt{A^2-B^2}<0$ $i.e.$  $\gamma/J>2$ then we are in PT-broken regime and the sinusoidal functions in eq. (\ref{6}) and ($\ref{7}$) change to hyperbolic functions, and periodicity of $S_{B}$ and $S_{n}$ is lost.

We now derive explicit expressions for the correlations for a typical TLS. For simplification of analytic expressions we choose $\vec{A}=J \  \hat{x}$, and $\vec{B} = \frac{\gamma}{2}  \ \hat{z}$. Then the non-Hermitian Hamiltonian $H_{PT} \ = J \sigma_{x} - i \frac{\gamma}{2} \ \sigma_{z}$ has two exceptional points at $\gamma = \pm 2 J$. It should be  noted that work along similar line has been done in \cite{Anant}. However there is a subtle difference that should be pointed out. Since a PT-symmetric Hamiltonian $H_{PT}$ is connected with a Hermitian Hamiltonian $\hat{h}$ via similarity transformation $P$ as: $ H_{PT}  = P \  \hat{h}  \ P^{-1}~$. For the Hermitian Hamiltonian $\hat{h}= \sigma_{x}$ the form of the similarity transformation  used in \cite{Anant} is given by $P= e^{\theta \sigma_{y}}$, where $\theta = (1/2) \tanh^{-1}(\sin \alpha) $. This transformation is such that it keeps the eigenspectrum fixed. However, here the similarity transformation connecting the $H_{PT}$ with the Hermitian counterpart: $\hat{h} = \frac{J}{2}  \sqrt{4-\frac{\gamma ^2}{J^2}}  \ \sigma_{x}$ is $P= e^{\theta \sigma_{y}}$, where $\theta = (1/2) \coth^{-1} (2J/\gamma)$, hence the eigenspectrum is tunable leading to the form of $H_{PT}= J \sigma_{x} - i \frac{\gamma}{2} \ \sigma_{z}$. We can now write  each components of the SOE ($v(t)$) which are  connected with the components of $\vec{S(t)}$ with the following relations:

\begin{eqnarray}
v_{1}(t) &=& \sqrt{ J^{2} \ ( \ 1 - 4  \ S_{x}^{2}(t) \ ) },
\label{3a} 
\end{eqnarray}
\begin{eqnarray}
v_{2}(t) &=& \sqrt{\dfrac{\gamma^{2}}{4} \ ( \ 1  - 4  \  S_{z}^{2}(t) \ ) },
\label{4a} 
\end{eqnarray}
\begin{eqnarray}
v_{3}(t) &=& \sqrt{2 J \gamma  \ S_{n}(t)}, 
\label{5a} 
\end{eqnarray}

 where $n$ denotes  the direction $\hat{n}= \hat{x} \times \hat{z}= -\hat{y}$. Utilizing the analytic expressions in the eqs. (\ref{3a}-\ref{5a}) above we can write the maximized SOE, $v^{max}=1+ \gamma/2$ given $J=1$. On the other hand minimum value of SOE, $v^{min}= 1-\gamma/2$ in the PT-symmetric phase and identically zero in the PT-broken phase.

\section{LGI and PT-symmetry \label{secIII}}

In this section we analyse the LG parameter $K_{3}$ across the exceptional point. We consider initial state as $\ket{\psi}= [\ \cos \frac{\theta}{2} e^{i \phi}, \sin \frac{\theta}{2} ]^{T} $ such that $\rho_{0} = \ket{\psi}\bra{\psi}$ and measurement operator of the form $Q= \vec{n} \cdot \sigma$ with $\vec{n}= ( \ \sin \theta_{m} \ \cos \phi_{m},\sin \theta_{m} \ \sin \phi_{m}, \cos \theta_{m} \  
)$. We also assume that $t_{1}=0$. Therefore, we have two time intervals namely, $t_{2}-t_{1}$ and $t_{3}-t_{2}$ when a state evolves, and measurements are made at three time instants $t_{1},t_{2}$ and $t_{3}$. We then numerically optimize the expression of $K_{3}$ over the complete parameter space comprising of six parameters $\{ \ \theta, \phi, \theta_{m},\phi_{m}, t_{2}, t_{3} \ \}$, for a fixed value of ratio $\gamma/J$ as illustrated in FIG. \ref{fig1} (Middle).


We find that for any non-zero value of $\gamma/J$, the maximum accessible value of $K_{3}$ is always greater than L\"{u}ders bound of $3/2$. Moreover, as we move across the exceptional point $i.e.$ $\gamma/J>2$, it is always possible to access the algebraic maximum of $K_{3}$. We now focus on the initial state and measurement operator dictating the optimization of the temporal correlations leading to this maximization of the LG parameter $K_{3}$. We note that the parameters $\phi$ and $\phi_{m}$ are $3 \pi/2$ and $\pi/2$ respectively, for all the maximized values of $K_{3}$ for different $\gamma$ values. This corresponds to initial state $\ket{\psi}$ and measurement direction $\vec{n}$ lying in the $\hat{y}-\hat{z}$ plane, in the optimized parameter space. This fact implies that all the trajectories followed by the state upon evolution that maximizes $K_{3}$ lie on the geodesic in the  $\hat{y}-\hat{z}$ plane on the Bloch sphere.  \\

{\underline{ PT-symmetric regime:-}} In this regime $\gamma /J < 2$, implying all the trajectories on the Bloch sphere are periodic in nature. The measurement operator $Q$ represented with $\vec{n}$ can be kept fixed to optimize $K_{3}$. As shown in the inset of FIG. \ref{fig1}, $\theta_{m}= \pi / 2$ $i.e.$ $Q= \sigma_{y}$. Fixing the measurement operator establishes one-to-one correspondence between the choice of the initial state $\ket{\psi}$ dictated by $\theta$ (red curve in the inset) and maximum value of $K_{3}$. Moreover, as we approach the exceptional point the initial state converges to one of the eigenstates of the Hamiltonian $H_{PT}$ $i.e.$ $\gamma/J \rightarrow 2$ implies $\ket{\psi} \rightarrow [1/\sqrt{2},i\sqrt{2}]^{T}$, which is an eigenstate of $H_{PT}$ when $\gamma/J \rightarrow 2$.  \\

{\underline{ PT-broken regime:-}}This regime corresponds to $\gamma/J>2$ and the periodicity of the trajectories followed by Bloch vector is lost. The two right eigenstates of the Hamiltonian $H_{PT}$ in this case corresponds to two fixed points of the dynamics. On the Bloch sphere one eigenstate acts as source and other eigenstate plays the role of sink. We find that in this regime it is always possible to access the algebraic maximum of $K_{3}$ (as shown in FIG. \ref{fig1}). We notice that initial state $\ket{\psi}$ and measurement operator $Q$, conspire together to bring out extreme temporal correlations resulting in $K_{3} \rightarrow 3$. In this regime $\theta = \theta_{m}$, implying that initial state Bloch vector $\vec{S}(0)$ and measurement operator direction $\vec{n}$ are always orthogonal to each other $i.e.$ initial state $\ket{\psi}$ is an eigenstate of the measurement operator $Q$. In what follows we explain how the algebraic maximum of $K_{3}$ occurs and its relation with the time dependent SOE especially in the PT-broken phase as $K_{3}\rightarrow 3$ is always possible in this phase.

\subsection{Algebraic bound and time dependent SOE}

 The algebraic bound of $K_{3}$ is special in the sense that it imposes stringent conditions on the joint probabilities and can only occur when $C_{12}=1$, $C_{12}=1$, and $C_{13}=-1$. In the temporal correlation $C_{12}=P_{12}(\uparrow,\uparrow) + P_{12}(\downarrow,\downarrow)- P_{12}(\uparrow,\downarrow)- P_{12}(\downarrow,\uparrow)$, the joint probabilities ought to be $P_{12}(\uparrow,\downarrow)= P_{12}(\downarrow,\uparrow)=0$. Important to note that ($\uparrow$, $\downarrow$) represent measurement operator eigenstates. Similarly,  $P_{23}(\uparrow,\downarrow), P_{23}(\downarrow,\uparrow)$ would also be identically zero in the correlation $C_{23}$. Finally, correlation $C_{13}$ must have $P_{13}(\uparrow,\uparrow) = P_{13}(\downarrow,\downarrow)=0$ in order to be equal to $-1$. A typical experiment resulting in the LG parameter $K_{3}$ approaching algebraic maximum is illustrated in the FIG. \ref{fig3}. We assume $t_{1}=0$ and equal time spacing $i.e.$ $t_{2}-t_{1}= t_{3}-t_{2}= \Delta t$. Subsequently, the evolution from time $t_{1}$ to $t_{2}$ is slow and state $\ket{\psi}$ remain very close to itself at time $t_{2}$. Therefore, we have $P_{12}(\uparrow,\uparrow) \approx 1$ and $P_{12}(\downarrow,\downarrow) \approx 0$ ( FIG. \ref{fig3} (a)), which implies $C_{12} \approx 1$. Similarly, the joint probabilities $P_{23}(\uparrow,\uparrow) \approx 1$ and $P_{23}(\downarrow,\downarrow) \approx 0$, owing to the fact that time evolved state is a function of time interval only and a measurement is performed at time $t_{2}$. Finally, in order to calculate $C_{13}$ the time interval is doubled is $2 \ t$ now. However, in this case $P_{13}(\uparrow,\downarrow) \approx 1$ and $P_{13}(\downarrow,\uparrow) \approx 0$, which results in $C_{13} \approx -1$.  
 
 \begin{figure}[htb!]
    \centering
    \hspace{-1em}
\includegraphics[width=7.0cm,height=5.5cm]{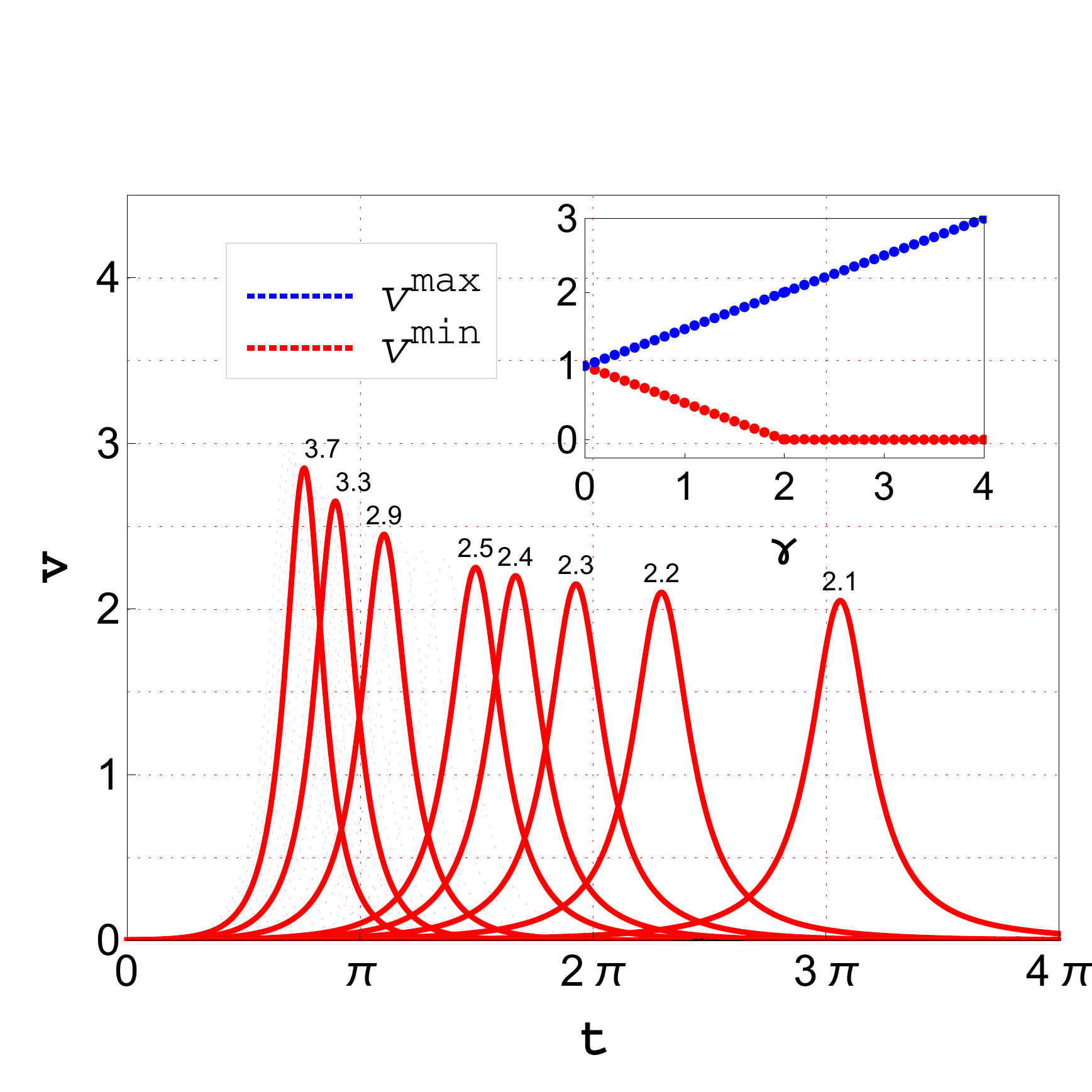}
\caption{SOE $v(t)$ varying in time when the PT-symmetry is broken $i.e.$ $\gamma>2$. Each curve corresponds to SOE for a different value of $\gamma$, indexed on the top of each curve. Inset shows the maximum and minimum values of speed denoted as $v^{max}=1+ \gamma/2$ and $v^{min}$ (equals to $1-\gamma/2$ in PT-symmetric phase and zero in the PT-broken phase) respectively in both PT-symmetric and broken regime.} 
\label{fig2}
\end{figure}

\begin{figure}[b!]
\centering
\includegraphics[width=7.0cm, height=5.5cm]{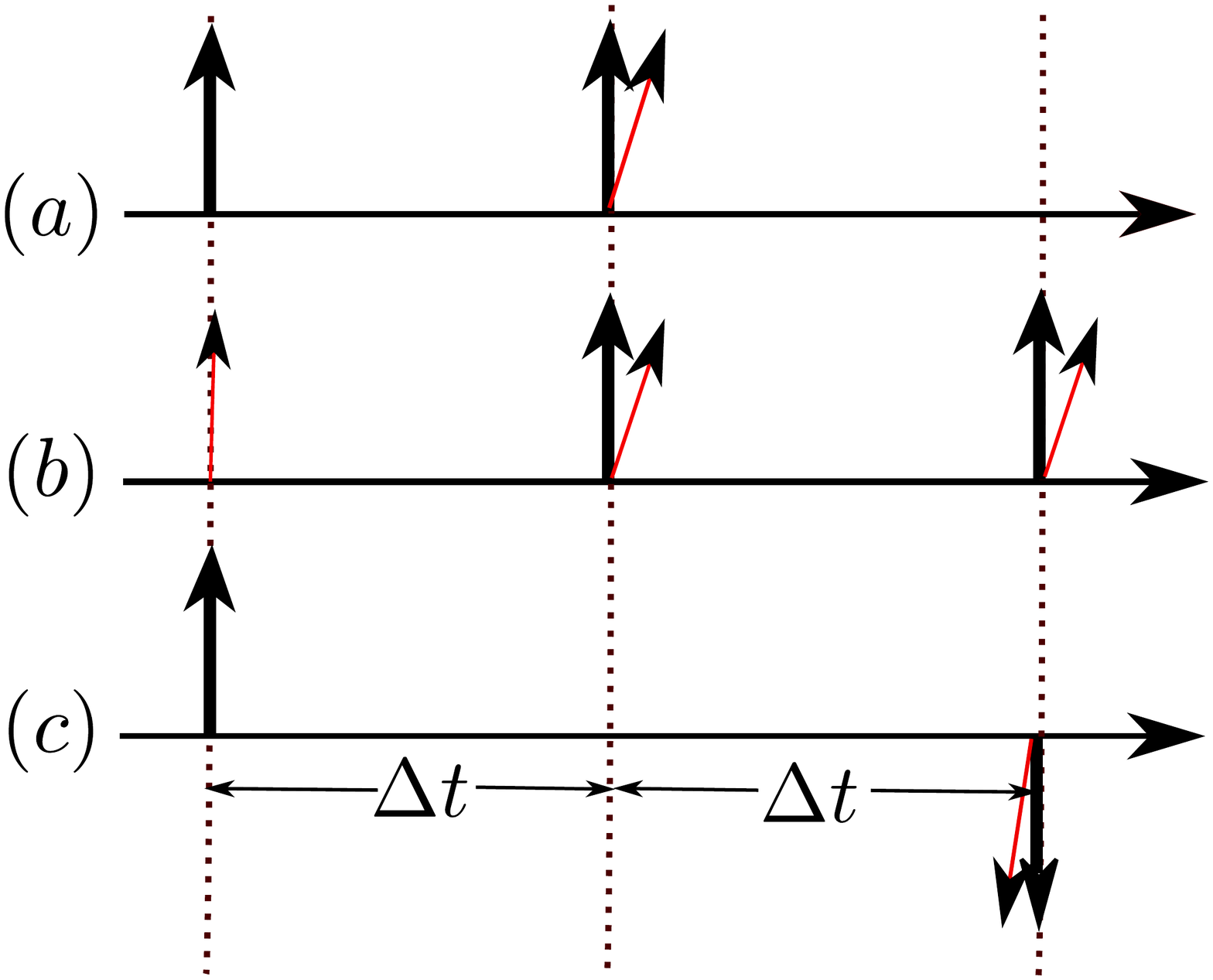}
\caption{ A schematic illustrating dynamical process corresponding to the case $K_{3} \rightarrow 3$. Correlations $C_{12}$, $C_{23}$ and $C_{13}$ are  calculated via the evolution  and measurements shown in (a), (b) and (c) respectively. }
\label{fig3}
\end{figure}

 Such extreme temporal correlations for a TLS have already been shown to connected with the change in the SOE of the Bloch vector in case of PT-symmetric dynamics \cite{Anant}. Therefore, following the same line of thought we explore how change in the SOE of the Bloch vector is responsible for the algebraic bound of 3 in the PT-broken regime. Since we know in this regime the initial state $\ket{\psi}$ for optimized $K_{3}$ value is an eigenstate of the measurement operator $Q$ (see FIG. \ref{fig1}). More explicitly initial state $\ket{\psi}= [\ \cos  (\theta_{m}/2) ,i \ \sin (\theta_{m}/2) ]^{T}= \ \uparrow $. Recalling there are two eigenstates of the Hamiltonian $H_{PT}$, which act as source (denote as $\ket{\uparrow}_{source}= \ket{+}$)  and sink (denote as $\ket{\downarrow}_{sink}=\ket{-}$)  for the dynamics in this regime. Interestingly, for optimized value of $K_{3}$ the source eigenstate is equal to initial state $i.e.$ $\ket{\uparrow}_{source} \approx \ket{\psi} = \ \uparrow$. It is reasonable to start with the initial state which is acting as a source, as once the state reaches the sink state all the dynamics ceases. It is clear now that in order to obtain $C_{12} \rightarrow 1$, $C_{23} \rightarrow 1$ and $C_{13} \rightarrow -1$ the evolution of the initial state $\ket{\psi} \approx \ket{\uparrow}_{source} $,  should be such that in the time interval $t$ it evolves very slowly to remain close to  state $\ket{\uparrow}_{source}$ and in the time interval $2 \Delta t$ it should flip close to its orthogonal state and then eventually evolves to the sink state $\ket{\downarrow}_{sink}$, in order to approach algebraic bond of $K_{3}$. We find that for optimized set of parameters the speeds of evolution throughout the PT-broken regime follow this trend as illustrated in FIG. \ref{fig2}.

\begin{widetext}
 
\begin{figure}[htb!]
    \centering
    \vspace{1em}
\includegraphics[width=4.0cm,height=4.5cm]{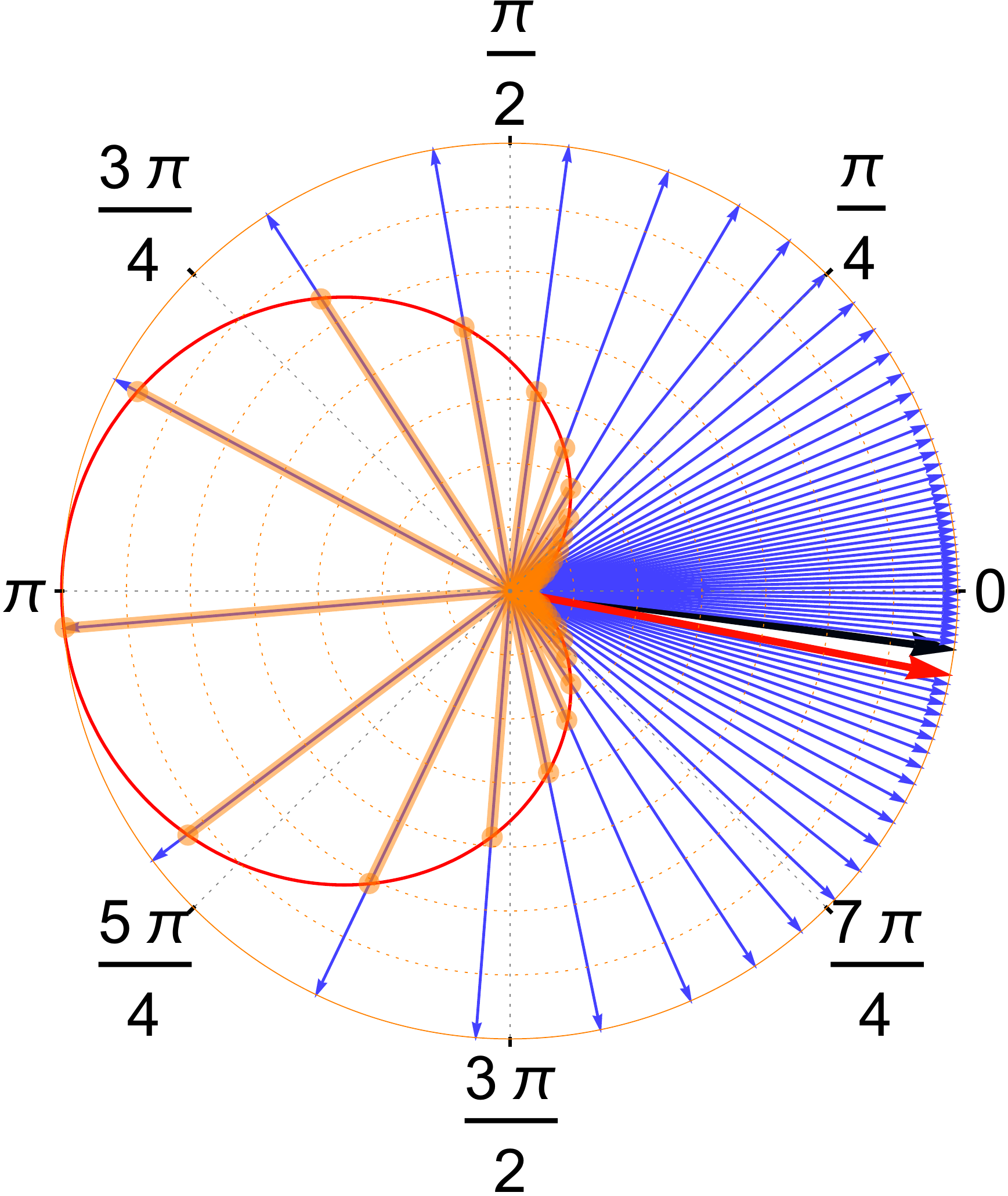} \hspace{1em}
\includegraphics[width=4.0cm,height=4.5cm]{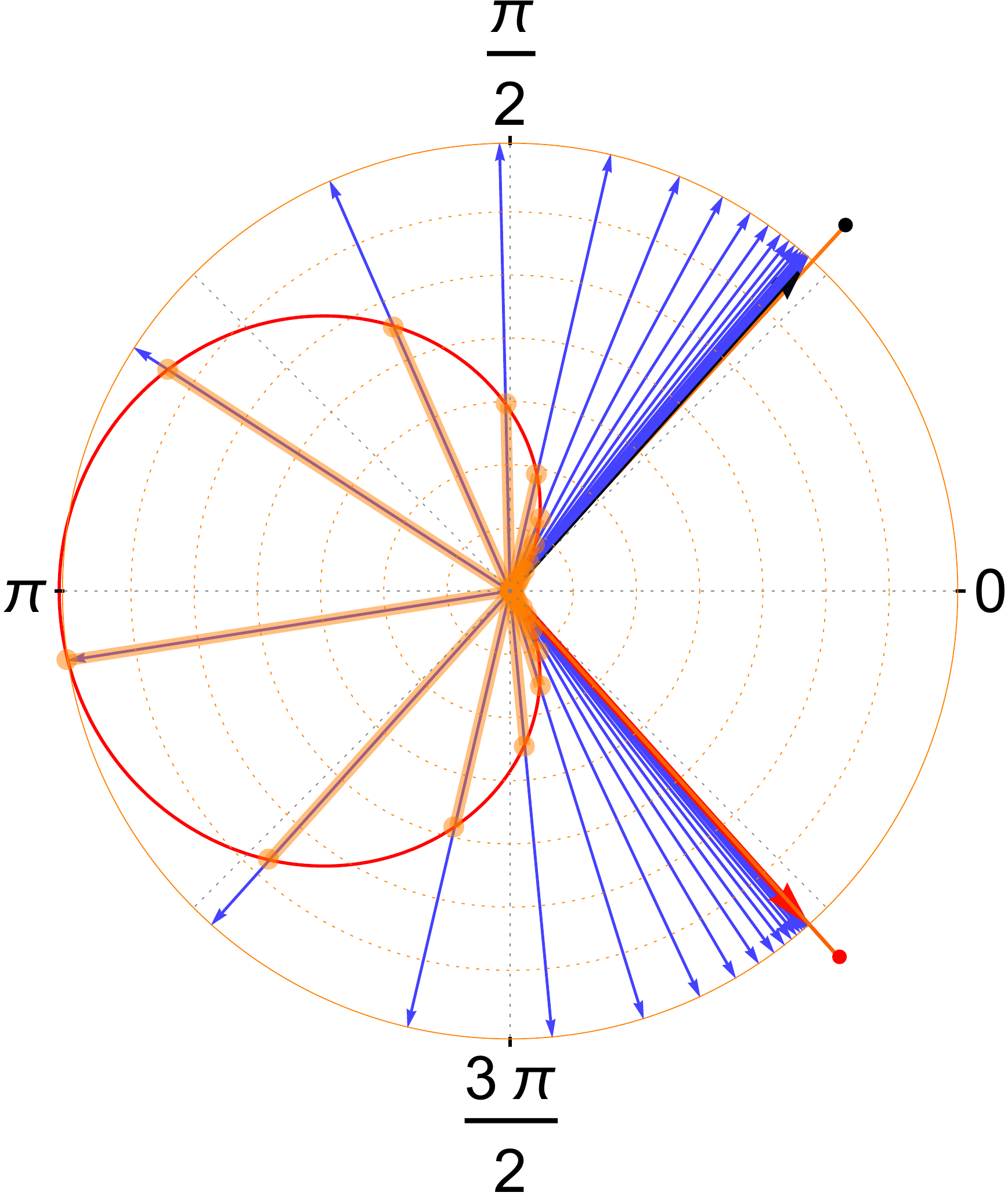} \hspace{1em}
\includegraphics[width=4.0cm,height=4.5cm]{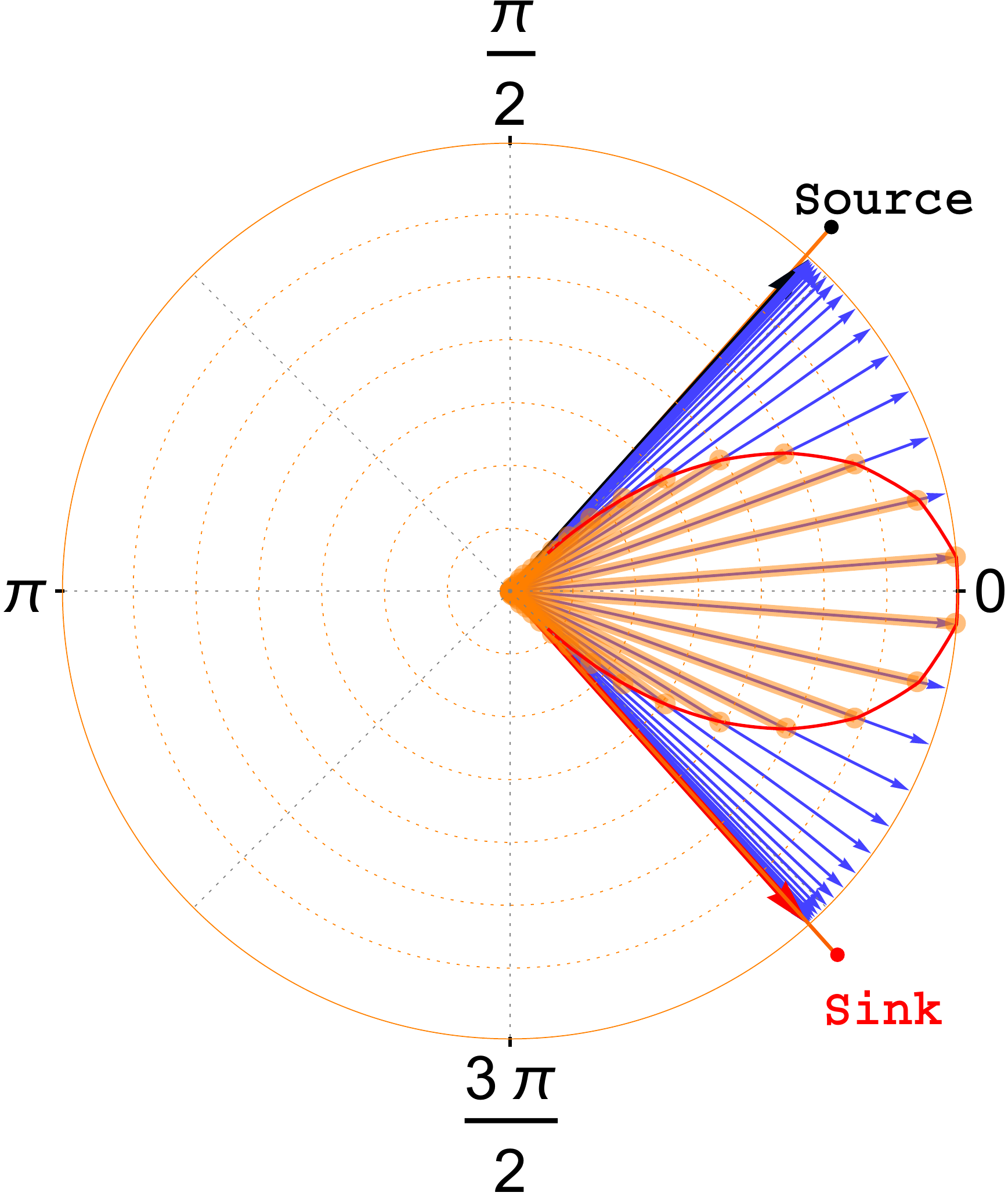} 
\caption{State evolution on $\hat{y}-\hat{z}$ plane geodesic with equal time interval spacing, for both the PT-symmetric (left) and PT-broken (middle \& right) phases. Here angles are defined by $\pi/2- \theta$ $i.e.$ zero corresponds to $\hat{y}$ axis and $\pi/2$ corresponds to $\hat{z}$ axis. Left polar plot corresponds to $\gamma/J=1.9$ which is PT-symmetric phase and at time $t_{1}=0$ the initial state $\ket{\psi}$ is represented with the black arrow and final state is illustrated with the red arrow. Middle polar plot corresponds to $\gamma/J=3$. In both the cases  $J=1$, and the total time interval for evolution in both the cases is same. The coefficient $ S_{n}(0) \ [ A+2 B S_{n}(0) ]<0$ in this case. Right polar plot is the evolution of the initial state chosen on the right hand side of the fixed point and evolves to the sink state fixed point owing to the coefficient $ S_{n}(0) \ [ A+2 B S_{n}(0) ]>0$, where $A=J=1$ and $B= \gamma/2= 3/2$. Also, plotted are the respective SOEs normalized as $v(t)/v^{max}$ for each case  and are denoted with orange lines and orange dots.} 
\label{fig5}
\end{figure}

\end{widetext}

In order to illustrate the above points we consider explicit example where $K_{3} \rightarrow 3$ in the PT-symmetric regime and PT broken regime. As mentioned earlier the trajectories followed by initial state lies on the $\hat{y}-\hat{z}$ plane geodesic of the Bloch sphere. In particular, we consider two different non-Hermiticity parameter (i) $\gamma/J=1.9$ $i.e.$ PT-symmetric regime and (ii) $\gamma/J=3$ which is deep PT-broken regime. It is clear from the FIG. \ref{fig5} (left \& middle) that in  both the cases the the maximum speed $v^{max}$ is achieved by the state when the $S_{B} \rightarrow 0$ in time also implying $S_{n} \rightarrow 1/2$. Moreover, in both cases the initial state $\ket{\psi}$ always evolves to its orthogonal state by choosing to evolve on the longer geodesic path. These two facts together confirm the dynamical process mentioned in the FIG. \ref{fig3}, leading to $K_{3} \rightarrow 3$. Important to note that the initial state choice $\ket{\psi}\approx \ket{\uparrow}_{source} $ is very sensitive to the neighbourhood of the $\ket{\uparrow}_{source}$ state. As shown in  FIG. \ref{fig5} (right) that choosing the initial state right to the fixed point  $\ket{\uparrow}_{source} $ would lead to evolution in the opposite direction compared to choosing the initial state on the left of the state  $\ket{\uparrow}_{source} $. Such evolution can never lead to  $K_{3} \rightarrow 3$.

\section{Simulating non-Hermitian dynamics with a three level open quantum system \label{secIV}}

In this section we first introduce a 3-level system which we exploit here to simulate the non-Hermitian dynamics of TLS as discussed so far. Then we will show how post-selecting the three level dynamics we can obtain an effective TLS dynamics which mimic the the non-Hermitian dynamics. Lets begin with the Lindbladian equation:

\begin{equation}
\dot{\rho_{3}} = \mathcal{L}(\rho_{3}) = -i[H, \rho_{3}] + \sum_{\alpha > 0} \gamma_\alpha \Big( L_\alpha \rho_{3} L^{\dagger}_\alpha - \frac{1}{2}\{ L^{\dagger}_\alpha  L_\alpha, \rho_{3} \}\Big)
\label{E6}
\end{equation}

We will now define a new operator $H_{eff}$, which is a non-Hermitian operator and would mimic the dynamics corresponding to $H_{PT}$ as follows:

\begin{equation}
H_{eff} = H - i\sum_\alpha \frac{\gamma_\alpha}{2}L^{\dagger}_\alpha L_\alpha 
\label{E7}
\end{equation}

With this new definition the Lindbladian equation now reduced to:

\begin{equation}
\frac{d \rho_{3}}{d t } =  -i (H_{eff} \rho - \rho H^{\dagger}_{eff}) +\sum_\alpha \gamma_\alpha L_\alpha \rho L^{\dagger}_\alpha 
\label{E8}
\end{equation}

The last term in eq. (\ref{E8}) is the quantum jump term between states in the $\rho$. Effective removal of this term would allow to simulate coherent non-unitary dynamics and can be used to recreate $H_{PT}$ dynamics.
Consider a 3-level atom coupled to cavity environment labeled as: $\vert f \rangle, \vert e \rangle, \vert g \rangle$. The levels are not equally separated ( $E_f-E_e \neq E_e- E_g$). Using these energy levels as the basis we define the following Hamiltonian $H$ and one dissipator $L_{1}$:

\begin{equation}
H = \begin{pmatrix}
0 & J & 0\\
J & 0 & 0\\
0 & 0 & -\epsilon_g 
\end{pmatrix}; 
L_{1} = \vert g \rangle \langle f \vert = \begin{pmatrix}
0 & 0 & 0\\
0 & 0 & 0\\
1 & 0 & 0 
\end{pmatrix}.
\label{E10}
\end{equation}

where  dissipator $\L_{1}$ has coefficient $\gamma_{1}$. It is to be noted that:

\begin{equation}
H- i \frac{\gamma_{1}}{2} \ L_{1}^{\dagger} L_{1} +  i \frac{\gamma_{1}}{4}  \begin{pmatrix}
1 & 0 & 0\\
0 & 1 & 0\\
0 & 0 & 0 
\end{pmatrix} = \begin{pmatrix}
-i \frac{\gamma_{1}}{4} & J & 0\\
J &  i \frac{\gamma_{1}}{4} & 0\\
0 & 0 & -\epsilon_g 
\end{pmatrix} .
\label{E11}
\end{equation}

Comparing the $2 \times 2$ block of the resultant matrix in eq. (\ref{E11}), with $H_{PT}$ we find that $\gamma_{1} = 2 \gamma$. Using  eq. (\ref{E10}) we can re-write the eqn. (\ref{E8}) in the matrix form with the assumption $J=1$ as:

\begin{widetext}
\begin{equation}
\frac{d \rho_{3}}{dt} =\frac{d}{dt}\begin{pmatrix}
 \rho_{ff} & \rho_{fe} & \rho_{fg} \\
\rho_{ef} & \rho_{ee} & \rho_{eg} \\
\rho_{gf} & \rho_{ge} & \rho_{gg} 
\end{pmatrix} = \begin{pmatrix}
i \ (\rho_{fe} - \rho_{ef}) - \gamma_1 \rho_{ff} & - \frac{\gamma_{1}}{2} \  \rho_{fe} + i \ (\rho_{ff} - \rho_{ee}) & -i (\epsilon_g  \ \rho_{fg} + \rho_{eg})-\frac{\gamma_1}{2}\rho_{fg} \\
-\frac{\gamma_{1}}{2} \  \rho_{ef} - i \ (\rho_{ff} - \rho_{ee}) & -i  \ (\rho_{fe}-\rho_{ef}) &  - i  \ (\rho_{fg} + \epsilon_g  \ \rho_{eg}) \\
i (\epsilon_g  \ \rho_{gf} + \rho_{ge})-\frac{\gamma_1}{2}\rho_{gf} & i (\rho_{gf} + \epsilon_g  \ \rho_{ge}) & \gamma_1 \ \rho_{ff}

\end{pmatrix}.
\label{E12}
\end{equation}
\end{widetext}

The dynamical equation (\ref{E12}) is exactly solvable, for the given setting. We find that time evolved elements of the density matrix $\rho_{3}$ namely $\rho_{fg}(t)$, $\rho_{eg}(t)$, $\rho_{gf}(t)$, and $\rho_{ge}(t)$ are zero at all times. Moreover, we find that the four elements $\rho_{ff}(t)$, $\rho_{fe}(t)$, $\rho_{ef}(t)$, and $\rho_{ee}(t)$ eventually die out in time and in the large time limit $ t \rightarrow \infty $, saturates to zero. In this limit though matrix element $ \rho_{gg}(t) \rightarrow 1 $. 

We now post select the $\vert f \rangle - \vert e \rangle$ block  of the three level time evolved density matrix $\rho(t)$ comprising of the elements $\rho_{ff}(t)$, $\rho_{fe}(t)$, $\rho_{ef}(t)$ and $\rho_{ee}(t)$. This $2 \times 2$ block (let's call $\rho_{2}(t)$)  $i.e.$

\begin{equation}
\rho_{2}(t)=  \begin{pmatrix}
\rho_{ff}(t) & \rho_{fe}(t) \\
\rho_{ef}(t) & \rho_{ee}(t) 
\end{pmatrix}.
\label{E13}
\end{equation}

Comparing $\rho_{2}(t)$ with the density matrix given in eq. (\ref{1}) we find that $\rho_{t}= \rho_{2}(t)/Tr[\rho_{2}(t)]$. Therefore, $\rho_{t}$ results on post-selecting the sub-ensemble corresponding to $\rho_{3}$.

\section{Discussion \label{secV}}

We illustrate  the existence of extreme temporal correlations  values $K_{3} \rightarrow 3$ in the PT-broken phase and close to the exceptional point in the PT-symmetric phase.  We establish the connection of the presence of the extreme temporal correlations with the SOE of the state and demonstrate the experiment that can access such correlations. Though the extreme values of the temporal correlations varies in the PT-symmetric and broken phase, both the phases share a common feature in terms of maximum value of SOE ($v^{max}= 1+ \gamma/2$), varying linearly in non-Hermitian parameter $\gamma$ independent of the phase. On the contrary, the minimum of SOE denoted as $v^{min}$ acts as the marker of  two distinct phases. While in the PT-symmetric phase $v^{min}= 1-\gamma/2$, in the PT-broken phase $v^{min}=0$.  It should be noted the for optimization of $K_{3}$ the full parameter space ($\{ \ \theta, \phi, \theta_{m},\phi_{m}, t_{2}, t_{3} \ \}$) possible has been considered. Consideration of all possible measurement angles for optimization of $K_{3}$ is important.

\begin{figure}[htb!]
   \centering
   \vspace{0em}
\includegraphics[width=4.275cm,height=5.0cm]{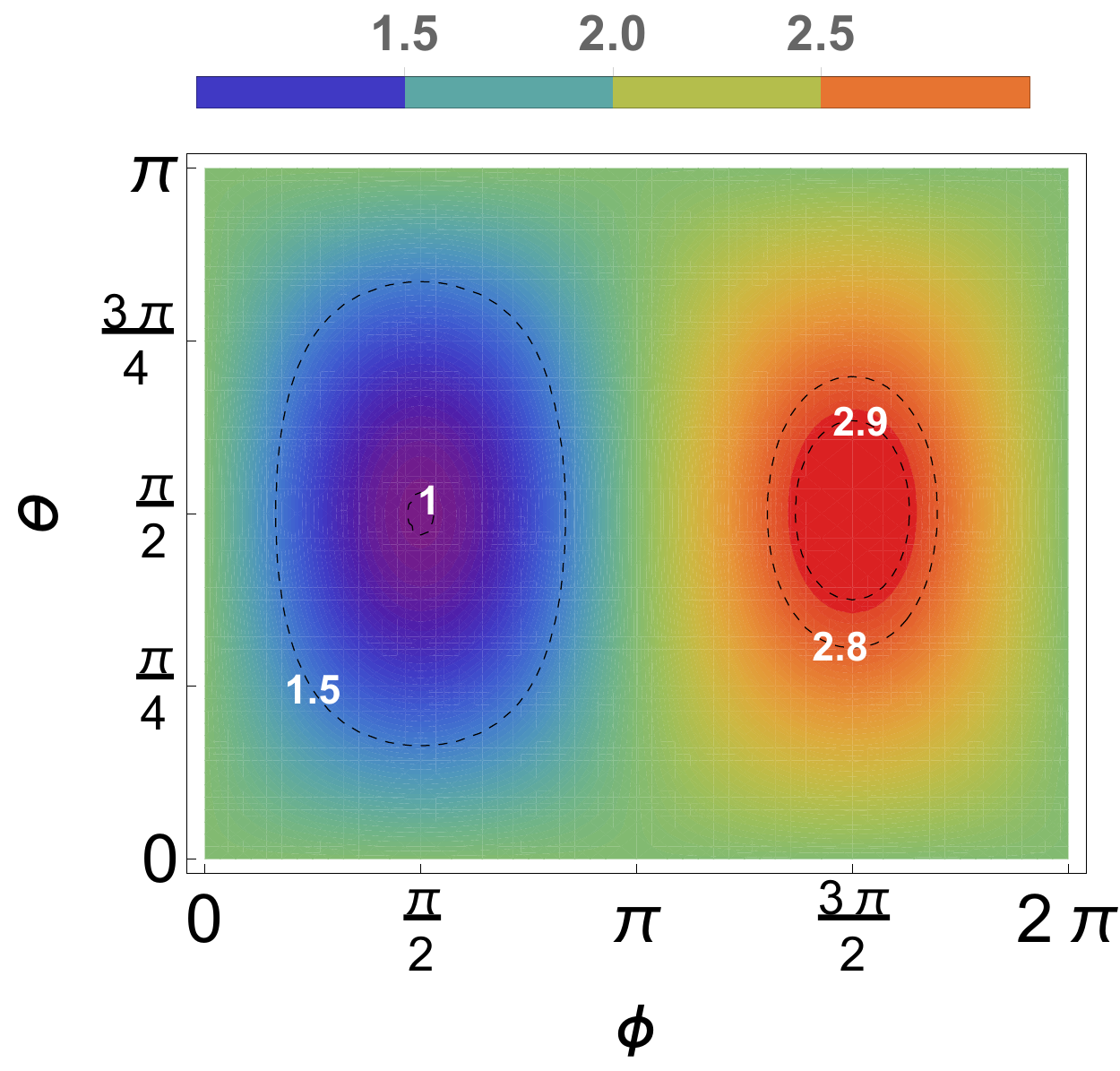} \hspace{-1em}
\includegraphics[width=4.275cm,height=5.0cm]{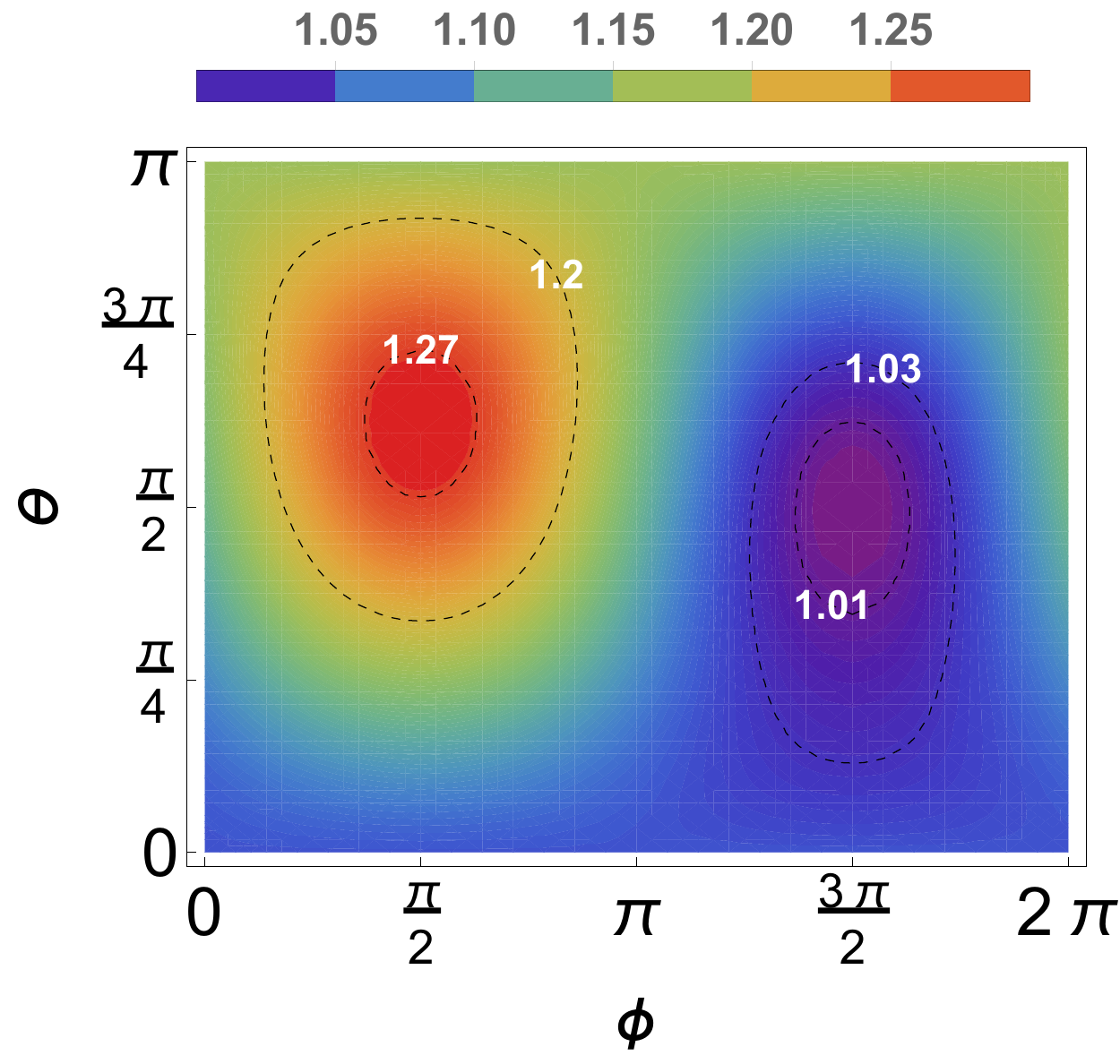}
\caption{Time optimized $K_{3}$ values in $\theta-\phi$ plane, where $\gamma=1.99$ \& $J=1$ $i.e.$ PT-symmetric phase (left). $K_{3}$ values in the PT-broken phase $i.e.$ $\gamma=2.01$ and $J=1$ (right).  } 
\label{fig6}
\end{figure}

For instance fixing the measurement angle denoted with $\vec{n}= \hat{y}$, would result in temporal correlations such that only in the PT-symmetric phase values close to algebraic bound can be found while in the PT-broken phase  L\"{u}ders bound is impossible to violate (FIG. \ref{fig6}).




 We finally  present the possible realization of non-Hermitian dynamics in three level atom coupled to a cavity environment. The subspace of this three level in which the non-Hermitian dynamics occurs can be parametrized by three parameters $r_{3}, \theta_{3},\phi_{3}$, which are functions of time. All possible dynamics of the TLS can be written in the following parametric form of the three level density matrix:

 \begin{equation}
     \ \rho_{3} =\begin{pmatrix}
 r_{3} (1+ \cos \theta_{3}) & e^{-i \phi_{3}} \ r_{3} \ \sin \theta_{3} & 0 \\
e^{i \phi_{3}} \ r_{3} \ \sin \theta_{3}& r_{3} (1- \cos \theta_{3}) & 0 \\
0 & 0 & 1- 2 r_{3},
\end{pmatrix}  
 \end{equation}

where $0 \leq r_{3} \leq 1/2$. The eigenvalues are $0, 2 r_{3}, 1-2 r_{3}$. Norm of the Bloch vector would be $\sqrt{1/3-2r_{3} + 4r_{3}^{2}}$. All the initial state chosen corresponds to $r_{3}=1/2$. The parameters $r_{3}, \theta_{3},\phi_{3}$ are function of $\gamma_{1}$, $J$ and time and intial state parameters (see appendix \ref{Three level dynamics}).

\section{Conclusion  \label{secVI}} 
In conclusion, we study temporal correlations quantified in terms of LG parameter $K_{3}$ across a PT-transition. We find that $K_{3}$ values being violated upto the algebraic bound can play a role of  marker which clearly distinguishes PT-symmetric side from the PT-broken side. We show that this extreme violation of LGI is directly related to the quantum speed limits of the evolution. Our finding clearly suggest that, the minimum SOE of the state in the full parameter space (initial state parameters and time) provides a clear signature of the PT-transition  and  is finite on the PT-symmetric side and identically zero on the PT-broken side hence can act as an order parameter.

\section{Acknowledgement} 
 It is a pleasure to thank Manas Kulkarni for initiating this collaboration. A.V.V. would like to thank the Council of Scientific and Industrial Research (CSIR), Govt. of India for financial support. S.D. would like to acknowledge the MATRICS grant (Grant No. MTR/ 2019/001 043) from the Science and Engineering Research Board (SERB) for funding.  Y.N.J. acknowledges funding from NSF Grant No. DMR-1054020.

\appendix

\section{SOE and spin components}

As mentioned in the \cite{Brody,Anant}, the SOE for a pure state $\ket{\psi}$, where evolution governed by the Hamiltonian $ H_{PT} \ = (\vec{A} - i \vec{B} ). \vec{ \sigma}$ can be written as:

\begin{equation}
v(t)^{2}= (\Delta \vec{A} \cdot \vec \sigma)^{2}+(\Delta \vec{B} \cdot \vec \sigma)^{2}  - i \left \langle [\vec{A} \cdot \vec \sigma,\vec{B} \cdot \vec \sigma]_{C} \right \rangle ,
\end{equation} 

where $(\Delta X \cdot \vec \sigma)^{2} = \left \langle  \ ( X.\vec{\sigma})^{2} \  \right \rangle - ( \ \left \langle X.\vec{\sigma} \right \rangle \ )^{2}$ is the variance of the operator, $\left \langle \   \right \rangle$ denote expectation value in the state $\ket{\psi}(t)$ and $ [ \ ]_{C} $ denotes commutator.  We can now re-write  $(\Delta X \cdot \vec \sigma)^{2}$ in terms of trace of the density matrix $\rho_{t} = \ket{\psi}(t)\bra{\psi}(t)$ given in eq. (\ref{1}) as:

\begin{equation}
(\Delta X \cdot \vec \sigma)^{2} = Tr[ \ \rho_{t}  (X.\vec{\sigma})^{2} \ ] - ( \  Tr[ \ \rho_{t} (X.\vec{\sigma})   \ ] \ )^{2}.
\end{equation} 

Using the definition of the density matrix $\rho_{t} =(1/2)~  \mathbb{I} + \vec{S}(t) \cdot \vec \sigma$ and the Hamiltonian $H_{PT} \ = \  (\vec{A} - i \vec{B} ). \vec{ \sigma} = \ J \sigma_{x} - i \frac{\gamma}{2} \ \sigma_{z}$, we can write:

\begin{equation}
(\Delta \vec{A} \cdot \vec \sigma)^{2} = J^{2} (\Delta \vec{ \sigma_{x}} \cdot \vec \sigma)^{2}  = J^{2} - 4 J^{2} \ S_{A}^{2}(t) 
\end{equation} 
\begin{equation}
(\Delta \vec{B} \cdot \vec \sigma)^{2} =\dfrac{ \gamma^{2}}{4} (\Delta \vec{ \sigma_{z}} \cdot \vec \sigma)^{2}  = \dfrac{ \gamma^{2}}{4} - \gamma^{2} \ S_{B}^{2}(t) \\
\end{equation} 
\begin{equation}
\left \langle [\vec{A} \cdot \vec \sigma,\vec{B} \cdot \vec \sigma]_{C} \right \rangle = J \gamma \left \langle [ \sigma_{x}, \sigma_{z}]_{C} \right \rangle =i 2 J \gamma S_{n}(t),
\end{equation} 

where $\hat{n} = \hat{A} \times \hat{B}$. We can now optimize the SOE $v(t)$ in its full parameter space. Since for maximization of $K_{3}$ we know that $S_{A}(t)=S_{A}(0)=0$ (see main text). This corresponds to $v_{1}(t)=J^{2}=1$. Moreover, the $v_{max}$ occurs at time $t_{max}$, which corresponds to $S_{B}(t_{max})=0$ and $S_{n}(t_{max})= 1/2$ leading to $v(t)= \sqrt{v_{1}(t)+ v_{2}(t) + v_{3}(t)}= \sqrt{1+ \gamma^{2}/4 + \gamma}= 1 + \gamma/2 $ (see inset of FIG. \ref{fig2}).


\section{Three level dynamics 
\label{Three level dynamics}}
The equation (\ref{E12}) is solvable exactly for all possible initial state in the $2 \times 2$ block of $\rho_{3}$. Following is the parametric form of the three level density matrix that includes all possible dynamics in the sub-space of TLS, which is undergoing non-Hermitian dynamics:

 \begin{equation}
     \ \rho_{3} =\begin{pmatrix}
 r_{3} (1+ \cos \theta_{3}) & e^{-i \phi_{3}} \ r_{3} \ \sin \theta_{3} & 0 \\
e^{i \phi_{3}} \ r_{3} \ \sin \theta_{3}& r_{3} (1- \cos \theta_{3}) & 0 \\
0 & 0 & 1- 2 r_{3},
\end{pmatrix}  
\label{B1}
 \end{equation}

where $0 \leq r_{3} \leq 1/2$. The eigenvalues are $0, 2 r_{3}, 1-2 r_{3}$. Norm of the Bloch vector would be $\sqrt{1/3-2r_{3} + 4r_{3}^{2}}$. All the initial state (at $i.e.   t=0$) chosen corresponds to $r_{3}=1/2$. This would corresponds to the identification $\theta_{3}= \theta$ and $\phi_{3}= \phi$. Moreover, the density matrix $\rho_{3}$ in eqn. (\ref{B1}) can also be written as the direct sum $\rho_{3}= 2 r_{3} \rho_{0} \oplus  (1-2 r_{3})$. 

The parameters $r_{3}, \theta_{3},\phi_{3}$ are function of $\gamma_{1}$, $J$ and time and initial state parameters. The solution of the equation (\ref{E12}) $i.e.$ time evolved density matrix $\rho_{3}(t)$ has one-to-one correspondence with the density matrix $\rho_{3}$ in the eqn. (\ref{B1}), with the following parametrization with $J=1$ and $\gamma_{1} = 2 \gamma$:

\begin{widetext}
\begin{equation*}
r_{3} = \frac{e^{- \gamma  t} (2 \gamma  s_{2} +\gamma  (\gamma -2 s_{2}) \cosh (t \omega )-\gamma  s_{3} \omega  \sinh (t \omega )-4)}{2 \left(\gamma ^2-4\right)},
\end{equation*}

\begin{equation*}
\theta_{3} =-\sec ^{-1}\left(\frac{2 e^{t \omega } \left(2 \omega  (\gamma  s_{2} - 2 )+\gamma  \omega  (\gamma -2 s_{2} ) \cosh (t \omega )-\gamma  \left(\gamma ^2-4\right) s_{3} \sinh (t \omega )\right)}{\left(\gamma ^2-4\right) \left(\gamma +2 s_{3} \left(e^{2 t \omega }-1\right)+e^{2 t \omega } (s_{3} \omega -\gamma ) + s_{3} \omega \right)}\right),
\end{equation*}

\begin{equation*}
\phi_{3}=i \left(t (\gamma +\omega )-\log \left(-\frac{A}{B}\right)\right),
\end{equation*}

where
\bea
   A &=&  i e^{t (\gamma +\omega )} \left(2 \omega  (\gamma  s_{2} - 2)+\gamma  \omega  (\gamma -2 s_{2} ) \cosh (t \omega )-\gamma  \left(\gamma ^2-4\right) \text{s3} \sinh (t \omega )\right)
                         \nn\\
                         && \times \sqrt{1-\frac{\left(\gamma ^2-4\right)^2 ((2 s_{2} -\gamma ) \sinh (t \omega )+ s_{3} \omega  \cosh (t \omega ))^2}{\left(2 \omega  (\gamma  s_{2} - 2)+\gamma  \omega  (\gamma -2 s_{2}) \cosh (t \omega )-\gamma  \left(\gamma ^2-4\right) s_{3} \sinh (t \omega )\right)^2}},
\label{E8}
\eea

\begin{equation*}
\text{B}=\omega  \left(\gamma  (\gamma  s_{2} - 2)+i \left(\gamma ^2-4\right) s_{1}\right)+2 \omega  (\gamma -2  s_{2}) \cosh (t \omega )-2 \left(\gamma ^2-4\right) s_{3} \sinh (t \omega )
\end{equation*}

\end{widetext}

and $\omega = \sqrt{-4 + \gamma^{2}}$, $s_{1} = \sin \theta \ \cos \phi$, $s_{2} = \sin \theta \ \sin \phi$ and $s_{3} = \cos \theta$.

Now for the illustration we present the solutions of the eqn. (\ref{E12}) for the initial state  of the form :

\begin{equation}
\rho(0) = \frac{1}{2}\begin{pmatrix}
1 & -i & 0\\
i & 1 & 0\\
0 & 0 & 0
\end{pmatrix}
\label{E14}
\end{equation} 

For the above initial density matrix (\ref{E14}) the solution of eqn. (\ref{E12}) is the following:

\begin{widetext}
\begin{equation*}
\rho_{ff}(t) = \frac{1}{2}e^{-\frac{\gamma_1 t}{2}}\Big( \frac{ 4 + \gamma_1 \cosh{(\frac{t}{2}\sqrt{\gamma_1^2-16})}-\sqrt{\gamma_1^2 -16} \sinh{(\frac{t}{2}\sqrt{\gamma_1^2 -16}})}{4 + \gamma_1 } \Big)
\end{equation*}
\begin{equation*}
\rho_{fe}(t) =  -\frac{i}{2}e^{-\frac{\gamma_1 t}{2}}\Big( \frac{ \gamma_1 + 4  \cosh{(\frac{t}{2}\sqrt{\gamma_1^2-16})}}{4 + \gamma_1 } \Big)
\end{equation*}
\begin{equation*}
\rho_{ef}(t) = \frac{i}{2}e^{-\frac{\gamma_1 t}{2}}\Big( \frac{ \gamma_1 + 4  \cosh{(\frac{t}{2}\sqrt{\gamma_1^2-16})}}{4 + \gamma_1 } \Big)
\end{equation*}
\begin{equation*}
\rho_{ee}(t) = \frac{1}{2}e^{-\frac{\gamma_1 t}{2}}\Big( \frac{ 4 + \gamma_1 \cosh{(\frac{t}{2}\sqrt{\gamma_1^2-16})}+\sqrt{\gamma_1^2 -16} \sinh{(\frac{t}{2}\sqrt{\gamma_1^2 -16}})}{4 + \gamma_1 } \Big)
\end{equation*}
\begin{equation*}
\rho_{gg}(t)  = 1-e^{-\frac{\gamma_1 t}{2}}\Big( \frac{ 4 +\gamma_1 \cosh{(\frac{t}{2}\sqrt{\gamma_1^2-16})}}{4 + \gamma_1 } \Big)
\end{equation*}
\begin{equation}
\rho_{fg}(t) = \rho_{eg}(t)  = \rho_{gf}(t) = \rho_{ge}(t) = 0
\end{equation}
\label{E15}
\end{widetext}

It is to be noted that the above analytical expression is matching with physical situation. For any non-zero $\gamma_1$, we see that as time goes to infinity all the other  elements of the $3\times 3$ $\rho$-matrix goes to zero except $\rho_{gg}(t) \longrightarrow 1$ which has to be the case for a spontaneous decaying atom.

The individual elements of $\rho^{N}_{2}$ (\ref{E13}) ( the density matrix for $\vert f \rangle - \vert e \rangle $ manifold) is given by the following:
\begin{widetext}
\begin{equation*}
\rho^N_{2,{ff}}(t) = \frac{1}{2}\Big( 1 - \frac{\sqrt{\gamma_1^2 -16} \sinh{(\frac{t}{2}\sqrt{\gamma_1^2 -16}})}{4 + \gamma_1 \cosh{(\frac{t}{2}\sqrt{\gamma_1^2-16})}} \Big), \quad \quad \quad \rho^N_{2,{fe}}(t) =  -\frac{2i}{\gamma_1 - \frac{\gamma_1^2 - 16}{\gamma_1 + 4 \cosh{(\frac{t}{2}\sqrt{\gamma_1^2 - 16})}}}
\end{equation*}
\begin{equation}
\rho^N_{2,ef}(t) =\frac{2i}{\gamma_1 - \frac{\gamma_1^2 - 16}{\gamma_1 + 4 \cosh{(\frac{t}{2}\sqrt{\gamma_1^2 - 16})}}}, \quad \quad \quad \rho^N_{2,ee}(t) = \frac{1}{2}\Big( 1 + \frac{\sqrt{\gamma_1^2 -16} \sinh{(\frac{t}{2}\sqrt{\gamma_1^2 -16}})}{4 + \gamma_1 \cosh{(\frac{t}{2}\sqrt{\gamma_1^2-16})}} \Big)
\end{equation}
\end{widetext}



\end{document}